\documentclass[floatfix,twocolumn,showpacs,preprintnumbers,amsmath,amssymb,prc,superscriptaddress]{revtex4} 
\usepackage{graphicx}% Includex figure files
\usepackage{dcolumn}% Align table columns on decimal point
\usepackage{bm}% bold math
\usepackage{color}
\begin{document}
%\linenumbers

\title{Beam-Energy Dependence of Directed Flow of Protons, Antiprotons and Pions in Au+Au Collisions}

\affiliation{AGH University of Science and Technology, Cracow, Poland}
\affiliation{Argonne National Laboratory, Argonne, Illinois 60439, USA}
\affiliation{University of Birmingham, Birmingham, United Kingdom}
\affiliation{Brookhaven National Laboratory, Upton, New York 11973, USA}
\affiliation{University of California, Berkeley, California 94720, USA}
\affiliation{University of California, Davis, California 95616, USA}
\affiliation{University of California, Los Angeles, California 90095, USA}
\affiliation{Universidade Estadual de Campinas, Sao Paulo, Brazil}
\affiliation{Central China Normal University (HZNU), Wuhan 430079, China}
\affiliation{University of Illinois at Chicago, Chicago, Illinois 60607, USA}
\affiliation{Cracow University of Technology, Cracow, Poland}
\affiliation{Creighton University, Omaha, Nebraska 68178, USA}
\affiliation{Czech Technical University in Prague, FNSPE, Prague, 115 19, Czech Republic}
\affiliation{Nuclear Physics Institute AS CR, 250 68 \v{R}e\v{z}/Prague, Czech Republic}
\affiliation{Frankfurt Institute for Advanced Studies FIAS, Germany}
\affiliation{Institute of Physics, Bhubaneswar 751005, India}
\affiliation{Indian Institute of Technology, Mumbai, India}
\affiliation{Indiana University, Bloomington, Indiana 47408, USA}
\affiliation{Alikhanov Institute for Theoretical and Experimental Physics, Moscow, Russia}
\affiliation{University of Jammu, Jammu 180001, India}
\affiliation{Joint Institute for Nuclear Research, Dubna, 141 980, Russia}
\affiliation{Kent State University, Kent, Ohio 44242, USA}
\affiliation{University of Kentucky, Lexington, Kentucky, 40506-0055, USA}
\affiliation{Korea Institute of Science and Technology Information, Daejeon, Korea}
\affiliation{Institute of Modern Physics, Lanzhou, China}
\affiliation{Lawrence Berkeley National Laboratory, Berkeley, California 94720, USA}
\affiliation{Massachusetts Institute of Technology, Cambridge, Massachusetts 02139-4307, USA}
\affiliation{Max-Planck-Institut f\"ur Physik, Munich, Germany}
\affiliation{Michigan State University, East Lansing, Michigan 48824, USA}
\affiliation{Moscow Engineering Physics Institute, Moscow Russia}
\affiliation{National Institute of Science Education and Research, Bhubaneswar 751005, India}
\affiliation{Ohio State University, Columbus, Ohio 43210, USA}
\affiliation{Old Dominion University, Norfolk, Virginia 23529, USA}
\affiliation{Institute of Nuclear Physics PAN, Cracow, Poland}
\affiliation{Panjab University, Chandigarh 160014, India}
\affiliation{Pennsylvania State University, University Park, Pennsylvania 16802, USA}
\affiliation{Institute of High Energy Physics, Protvino, Russia}
\affiliation{Purdue University, West Lafayette, Indiana 47907, USA}
\affiliation{Pusan National University, Pusan, Republic of Korea}
\affiliation{University of Rajasthan, Jaipur 302004, India}
\affiliation{Rice University, Houston, Texas 77251, USA}
\affiliation{University of Science and Technology of China, Hefei 230026, China}
\affiliation{Shandong University, Jinan, Shandong 250100, China}
\affiliation{Shanghai Institute of Applied Physics, Shanghai 201800, China}
\affiliation{SUBATECH, Nantes, France}
\affiliation{Temple University, Philadelphia, Pennsylvania 19122, USA}
\affiliation{Texas A\&M University, College Station, Texas 77843, USA}
\affiliation{University of Texas, Austin, Texas 78712, USA}
\affiliation{University of Houston, Houston, Texas 77204, USA}
\affiliation{Tsinghua University, Beijing 100084, China}
\affiliation{United States Naval Academy, Annapolis, Maryland, 21402, USA}
\affiliation{Valparaiso University, Valparaiso, Indiana 46383, USA}
\affiliation{Variable Energy Cyclotron Centre, Kolkata 700064, India}
\affiliation{Warsaw University of Technology, Warsaw, Poland}
\affiliation{University of Washington, Seattle, Washington 98195, USA}
\affiliation{Wayne State University, Detroit, Michigan 48201, USA}
\affiliation{Yale University, New Haven, Connecticut 06520, USA}
\affiliation{University of Zagreb, Zagreb, HR-10002, Croatia}

\author{L.~Adamczyk}\affiliation{AGH University of Science and Technology, Cracow, Poland}
\author{J.~K.~Adkins}\affiliation{University of Kentucky, Lexington, Kentucky, 40506-0055, USA}
\author{G.~Agakishiev}\affiliation{Joint Institute for Nuclear Research, Dubna, 141 980, Russia}
\author{M.~M.~Aggarwal}\affiliation{Panjab University, Chandigarh 160014, India}
\author{Z.~Ahammed}\affiliation{Variable Energy Cyclotron Centre, Kolkata 700064, India}
\author{I.~Alekseev}\affiliation{Alikhanov Institute for Theoretical and Experimental Physics, Moscow, Russia}
\author{J.~Alford}\affiliation{Kent State University, Kent, Ohio 44242, USA}
\author{C.~D.~Anson}\affiliation{Ohio State University, Columbus, Ohio 43210, USA}
\author{A.~Aparin}\affiliation{Joint Institute for Nuclear Research, Dubna, 141 980, Russia}
\author{D.~Arkhipkin}\affiliation{Brookhaven National Laboratory, Upton, New York 11973, USA}
% ...

\collaboration{STAR Collaboration}\noaffiliation

\begin{abstract}
Rapidity-odd directed flow ($v_1$) measurements for charged pions, protons and antiprotons near mid-rapidity ($y=0$) are reported in $\sqrt{s_{NN}} =$ 7.7, 11.5, 19.6, 27, 39, 62.4 and 200 GeV Au + Au collisions as recorded by the STAR detector at the Relativistic Heavy Ion Collider (RHIC).   At intermediate impact parameters, the proton and net-proton slope parameter $dv_1/dy|_{y=0}$ shows a minimum between 11.5 and 19.6 GeV. In addition, the net-proton $dv_1/dy|_{y=0}$ changes sign twice between 7.7 and 39 GeV. The proton and net-proton results qualitatively resemble predictions of a hydrodynamic model with a first-order phase transition from hadronic matter to deconfined matter, and differ from hadronic transport calculations.
\end{abstract}

\maketitle
%\begin{linenumbers}

%\realpagewiselinenumbers
%\setlength\linenumbersep{0.10cm}
% 0.10 is the correct number.

Lattice QCD calculations indicate that the transition from hadronic matter to a Quark Gluon Plasma (QGP) \cite{whitepapers} phase in gold ion collisions at the full energy of the Relativistic Heavy Ion Collider ($\sqrt{s_{NN}} \sim 200$ GeV) is a smooth crossover \cite{lattice1}, whereas at progressively lower beam energies, there is an increasing possibility to explore a first-order transition between these phases~\cite{ejiri,kapusta}. At even lower beam energies (how low remains unknown), the excited nuclear matter is expected to remain in a hadronic phase throughout the interaction \cite{whitepapers}. In the scenario of nuclear collisions at the optimum beam energy for a first-order phase transition, various models have predicted characteristic azimuthal anisotropy signals \cite{Rischke,Csernai,Brachmann,Stoecker}.

The first harmonic coefficient of the Fourier expansion of the final-state momentum-space azimuthal distribution relative to the reaction plane \cite{methods} is called directed flow $v_1$. The rapidity-even component $v_1^{\rm even}(y)$, attributed to event-by-event fluctuations in the initial state of the collisions, is unrelated to the reaction plane \cite{Luzum}. The $v_1^{\rm odd}(y)$ component is the traditional definition of $v_1$ as used for more than two decades \cite{Ollitrault1992,methods}, and is attributed to collective sidewards deflection of the particles. Both hydrodynamic \cite{Hydro} and nuclear transport \cite{urqmd} models indicate that $v_1(y)$ in the mid-rapidity region offers sensitivity to details of the expansion of the participant matter during the early collision stages \cite{prequilibrium,Heinz}. Hydrodynamic models predict a minimum in directed flow (e.g., a minimum in $dv_1/dy$) near mid-rapidity as a function of collision energy~\cite{Rischke,Stoecker}.  A three-fluid hydrodynamic calculation, with a first-order phase transition between hadronic matter and a quark gluon plasma, predicts a prominent minimum in directed flow of net baryons \cite{Stoecker2013} at a center-of-mass energy of about $\sqrt{s_{NN}} = 4$ GeV, and this minimum has been termed the ``softest point collapse"~\cite{Stoecker}.

The established convention assigns a positive sign to $v_1$ for nucleons detected near beam rapidity on whichever side of mid-rapidity has been arbitrarily defined as positive rapidity \cite{E895,NA49_v1,v1-62,GangThesis}. Predictions of hydrodynamic and transport models include $dv_1/dy$ with negative sign near mid-rapidity (where pions dominate), and such phenomena have been given names like ``anti-flow" \cite{Brachmann}, ``third flow component" \cite{Csernai} and ``wiggle" \cite{Wiggle,Herrmann,Stoecker}.  It has been argued that these are possible phase transition signatures, especially if observed for baryons~\cite{Stoecker}.  However, it is also possible to explain some qualitative features of a single sign reversal in $dv_1/dy$ in a purely hadronic picture \cite{Wiggle,Aihong-v1Model}. These alternative explanations imply that emission from a tilted disk-shaped source is similar for both pions and protons; both should show directed flow in the same direction close to mid-rapidity, and in the opposite direction in the region where spectator matter breaks up~\cite{Bozek}. 

We report measurements of directed flow in Au+Au collisions in the range $\sqrt{s_{NN}}=$ 7.7 to 200 GeV, based on data from STAR~\cite{starnim}, recorded in 2010 and 2011. The STAR Time Projection Chamber (TPC) \cite{startpc} performed charged particle tracking at mid-rapidity.  The centrality was determined from the number of charged particles in the pseudorapidity region $ |\eta|  < 0.5$.  Two Beam-Beam Counters~\cite{BBC,trigger} covering $3.3 < |\eta| < 5.0$ were used to reconstruct the first-harmonic event plane, as explained in Ref.~\cite{CuCuPaper}.  The BBC event-plane resolution is inadequate above 39 GeV, and therefore the STAR ZDC-SMD detectors were used at 62.4 and 200 GeV \cite{GangThesis,v1-62,v1-4systems,pidv1_200}.  

The analyzed data sets at 7.7, 11.5, 19.6, 27, 39, 62.4 and 200 GeV contain 3.8, 10.6, 19, 39, 96, 50 and 250 million minimum-bias-trigger events, respectively.  We require the primary vertex position of each event along the beam direction, $V_z$, to lie within 30 cm of the center of the detector for beam energies 39, 62.4 and 200 GeV, 40 cm for 27 and 19.6 GeV, 50 cm for 11.5 GeV and 70 cm for 7.7 GeV. Use of the same narrow or wide $|V_z|$ cut at all energies negligibly changes $v_1$, but a wider cut reduces statistical errors at the two lowest energies. Tracks were required to have transverse momenta $ p_T > 0.2 $ GeV$/c$, have a distance of closest approach to the primary vertex of less than 3 cm, have at least 15 space points in the main TPC acceptance $(|\eta| < 1.0)$, and have a ratio of the number of measured space points to the maximum possible number of space points greater than 0.52.  This last requirement prevents track splitting from causing a single particle to be counted twice.  Protons and antiprotons with $p_T$ between 0.4 and 2.0 GeV/$c$ and $\pi^\pm$ with $p_T > 0.2$ GeV$/c$ and $p$ up to 1.6 GeV/$c$ were identified based on energy loss in the TPC, and time-of-flight information from the TOF detector \cite{TOF}.  Intermediate-centrality proton and antiproton results presented in this study have been weighted, within the indicated acceptance, to correct for $p_T$-dependent inefficiency; these corrections are small, and are comparable to statistical uncertainties. For other particles, the presented results are uncorrected.   

Possible systematic uncertainties arising from non-flow, i.e., azimuthal correlations not related to the reaction plane orientation (arising from resonances, jets, strings, quantum statistics, and final-state interactions like Coulomb effects) are reduced due to the relatively large pseudorapidity gap between the STAR TPC and the BBC detectors~\cite{methods,CuCuPaper}. Directed flow measurements based on the BBC event plane, where the BBC east and west detectors ensure symmetry in rapidity acceptance, cancel biases from conservation of momentum in the basic correlation measurement, because the difference in $v_1$ between the rapidity hemispheres is used \cite{MomentumCons}. However, momentum conservation effects~\cite{MomentumCons} do contribute to systematic uncertainty in the event-plane resolution, and thereby in the resolution-corrected signal, at the level of less than 2\% \cite{simulations}.  A correction for weak decay feed-down is unnecessary for the particle species considered here and is neglected \cite{pidv1_200}.  The systematic uncertainty arising from particle misidentification and detector inefficiency is estimated by varying event and track cuts, and is $\sim 5$\%.  Simulations based on the UrQMD transport model \cite{urqmd} indicate that possible systematic effects due to the rapidity coverage of the event plane detectors is well within the total systematic uncertainty. The measured $v_1$ should be antisymmetric about mid-pseudorapidity within statistical uncertainties; previous studies suggest that the maximum forward-backward difference is a useful estimator of the systematic uncertainty~\cite{v1v2v4-200-long}.  Overall, total systematic uncertainties on $dv_1/dy$ are typically within 12\% in regions where this slope is not close to zero, and decrease slightly with increasing beam energy up to 39 GeV.  Specific point-by-point systematic errors are presented in the final two figures. 
 
%%%%%%%%%%%%%% Fig. 1 %%%%%%%%%%%%%%%%%%%%%%%%%%%%%
\begin{figure}[!htb]
\begin{center}
\includegraphics[scale=0.40]{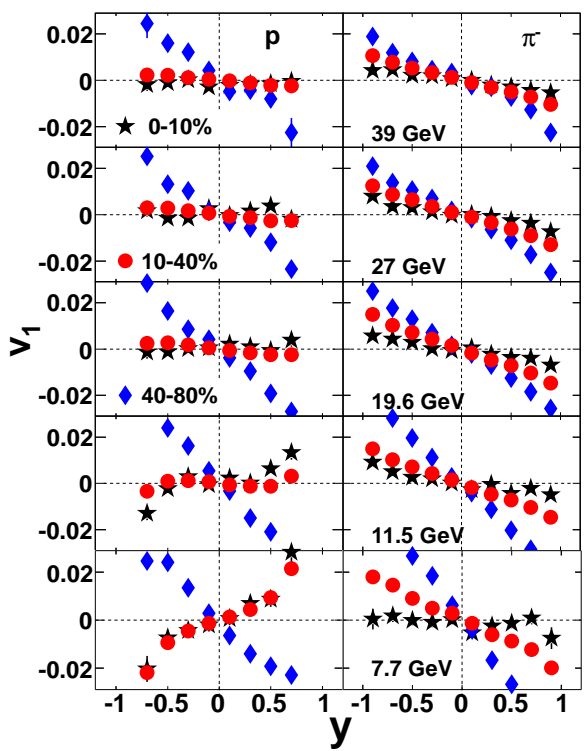}
\caption{ Directed flow for protons and $\pi^-$ versus rapidity for central (0-10\%), intermediate-centrality (10-40\%) and peripheral (40-80\%) Au+Au collisions at $\sqrt{s_{NN}}$= 39, 27, 19.6, 11.5 and 7.7 GeV.  The plotted errors are statistical only, and systematic uncertainties are explained in the text. }
\label{fig1}
\end{center}
\end{figure}
\begin{figure}[!htb]
\begin{center}
\includegraphics[scale=0.48]{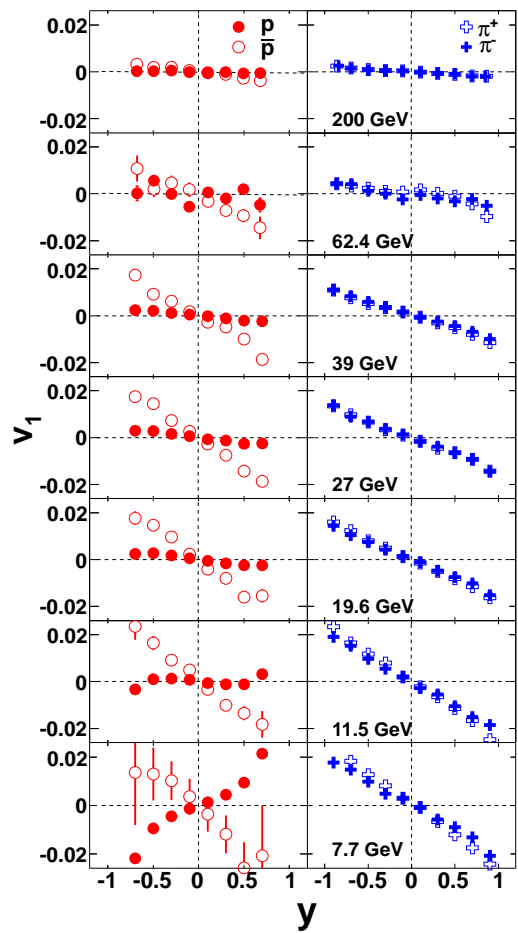}
\caption{ Proton and antiproton $v_1(y)$ (left panels) and $\pi^\pm$ $v_1(y)$ (right panels) for intermediate-centrality (10-40\%) Au+Au collisions at 200, 62.4, 39, 27, 19.6, 11.5 and 7.7 GeV.  The plotted errors are statistical only. }
\label{fig2}
\end{center}
\end{figure}
\begin{figure}[!htb]
\begin{center}
\includegraphics[scale=0.40]{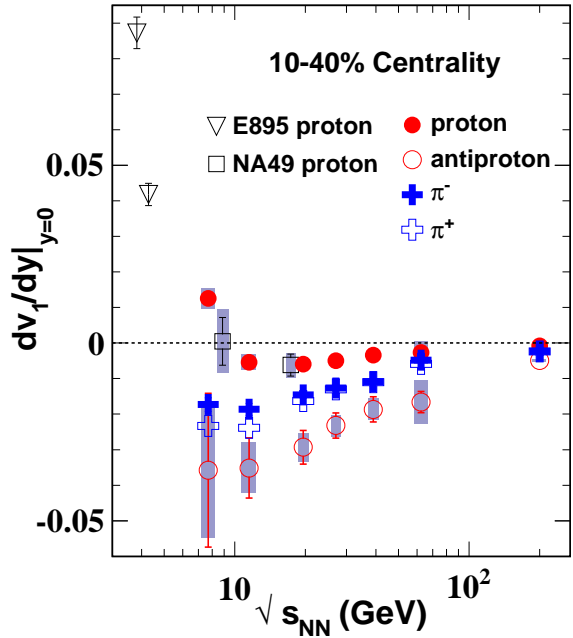}
\caption{ Directed flow slope ($dv_1/dy$) near mid-rapidity versus beam energy for intermediate-centrality Au+Au collisions.  The slopes for protons, antiprotons and $\pi^\pm$ are reported, along with measurements by prior experiments \cite{E895,NA49_v1} with comparable but not identical cuts.  Statistical errors (bars) and systematic errors (shaded) are shown separately. }
\label{fig3}
\end{center}
\end{figure}
\begin{figure}[!htb]
\begin{center}
\includegraphics[scale=0.40]{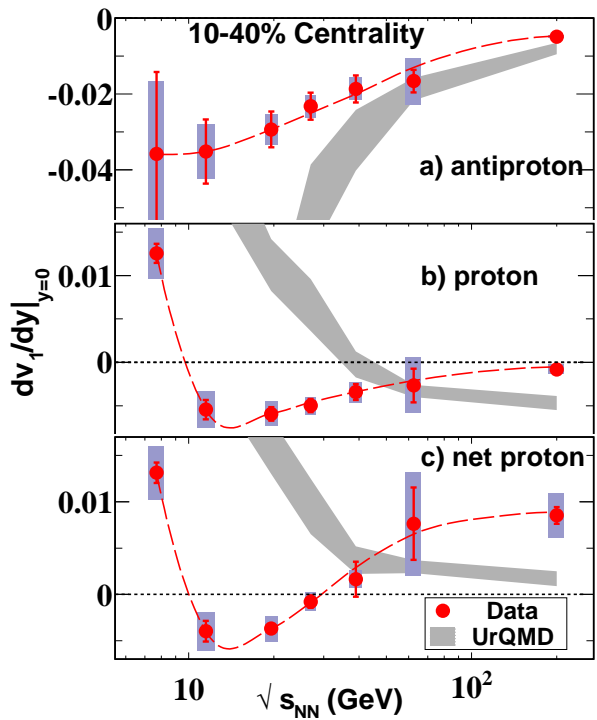}
\caption{ Directed flow slope ($dv_1/dy$) near mid-rapidity versus beam energy for intermediate-centrality Au+Au.  Panels (a), (b) and (c) report measurement for antiprotons, protons, and net protons, respectively, along with UrQMD calculations subject to the same cuts and fit conditions. Systematic uncertainties are shown as shaded bars. Dashed curves are a smooth fit to guide the eye. }
\label{fig4}
\end{center}
\end{figure}

%%%%%%%%%%%%%%%%%%%%%%%%%%%%%%%%%%%%%%%%%%%%%%%%%%%%
In Fig.~\ref{fig1}, $v_1(y)$ for protons and $\pi^-$ is presented for three centralities at $7.7-39$ GeV. Directed flow from STAR at 62.4 and 200 GeV has already been published \cite{v1-4systems,pidv1_200}.  A new analysis of later experimental runs with improved statistics is included in Figs.~\ref{fig2} through \ref{fig4}, and is consistent with the earlier measurements.  All data points in Fig.~\ref{fig1} are antisymmetric about mid-rapidity, verifying cancellation of the momentum conservation effect discussed earlier \cite{MomentumCons}. In intermediate and peripheral collisions, slopes of $v_1(y)$ near mid-rapidity for pions and protons are negative for all energies, except for protons at 7.7 GeV.  The NA49 collaboration \cite{NA49_v1} likewise has reported negative slopes at mid-rapidity for pions and protons at 17.3 GeV, with larger errors.  Furthermore, STAR has previously reported negative slopes at mid-rapidity for pions and protons at higher beam energies \cite{v1-4systems,pidv1_200}. 

These results cannot be explained by a baryon stopping picture~\cite{Wiggle}, which predicts a small slope for pions and an opposite slope for protons, contrary to the present observation of a large pion $v_1(y)$ slope that is not opposite to the proton $v_1(y)$ slope, except at 7.7 GeV.  Both protons and pions above 11.5 GeV have negative $dv_1/dy$ near mid-rapidity, which is consistent with predictions based on emission from a tilted source~\cite{Bozek}. Spectator shadowing also leads to negative $dv_1/dy$ for protons, with the most pronounced effect in peripheral collisions; however, its beam energy dependence has not been reported \cite{mimicQGP}.

%%%%%%%%%%%%%%%%%%%%%%%%%%% FIG 2 %%%%%%%%%%%%%%%%%%%%%%%%%%%
%%%%%%%%%%%%%%%%%%%%%%%%%%%%%%%%%%%%%%%%%%%%%%%%%%%%%%%%%%%%%

In Fig.~\ref{fig2}, $v_1(y)$ for protons, antiprotons and $\pi^\pm$ are presented for 10-40\% centrality Au+Au collisions at all of the studied beam energies.  We observe a large percentage difference between proton and antiproton $v_1$ at all seven energies. $v_1(y)$ is close for $\pi^+$ and $\pi^-$ at the higher energies, with minor  differences at 11.5 and 7.7 GeV. 

%%%%%%%%%%%%%%%%%%%%%%%%%%% FIG 3 %%%%%%%%%%%%%%%%%%%%%
%%%%%%%%%%%%%%%%%%%%%%%%%%%%%%%%%%%%%%%%%%%%%%%%%%%%%%%
Figure~\ref{fig3} presents $v_1(y)$ slope near mid-rapidity for protons, antiprotons and $\pi^\pm$ versus beam energy.  The slope is the linear term $F$ in a cubic fit, where $v_1(y) = F y + F_3 y^3$.  Figure~\ref{fig4}(a) duplicates the antiproton data, and Fig.~\ref{fig4}(b) shows the proton data in more detail; in both cases, UrQMD hadronic transport model~\cite{urqmd,Petersen2006} predictions are overlaid.  The fitted $F$ values are stable when binning and the $y$ range are varied (plotted slopes at and below 39 GeV are based on $-0.5 < y < 0.5$), and the systematic errors plotted in Figs.~\ref{fig3} and \ref{fig4} include uncertainties arising from the choice of fitting criteria.  At 62.4 and 200 GeV, a linear-only fit over $-1 < y < 1$ was used, but the systematic error covers the range of slopes from a cubic fit.

For intermediate-centrality collisions, the proton slope decreases with increasing energy and changes sign from positive to negative between 7.7 and 11.5 GeV, shows a minimum between 11.5 and 19.6 GeV, and remains small and negative up to 200 GeV, while the pion and antiproton slopes are negative at all measured energies. In contrast, there is no hint of the observed non-monotonic behavior for protons in the well-tested UrQMD model. Isse {\it et al.}, in a transport model study incorporating a momentum-dependent mean field, report qualitative reproduction \cite{JAM} of proton directed flow from E895 \cite{E895} and NA49 \cite{NA49_v1} (see Fig.~\ref{fig3}), but this model yields a positive $dv_1/dy$ at all beam energies studied ($\sqrt{s_{NN}}= 17.2$, 8.8 GeV and below).  

The energy dependence of proton $dv_1/dy$ involves an interplay between the directed flow of protons associated with baryon number transported from the initial beam rapidity to the vicinity of mid-rapidity, and the directed flow of protons from particle-antiparticle pairs produced near mid-rapidity.  The importance of the second mechanism increases strongly with beam energy. A means to distinguish between the two mechanisms would thus be informative.  We define the slope $F_{{\rm net\mbox{-}}p}$ based on expressing the rapidity dependence of directed flow for all protons as $[v_1(y)]_p = r(y) [v_1(y)]_{\bar{p}} + [1-r(y)]\, [v_1(y)]_{{\rm net\mbox{-}}p}$, where $r(y)$ is the observed rapidity dependence of the ratio of antiprotons to protons at each beam energy.  Corrections of $r(y)$ for reconstruction inefficiency and backgrounds have a negligible effect on $F_{{\rm net\mbox{-}}p}$ and have not been applied.  An interpretation of $F_{{\rm net\mbox{-}}p}$ is suggested by our observation that $v_1(y)$ is very similar for $\pi^+$ and $\pi^-$ (see Fig.~\ref{fig2}) and for $K^+$ and $K^-$ \cite{LongPaperBESv1}. Thus, we propose the use of antiproton directed flow as a proxy for the directed flow of produced protons, and propose that the net-proton slope $F_{{\rm net\mbox{-}}p}$ brings us a step closer to isolating the contributions from transported initial-state baryonic matter, as well as closer to the net-baryon hydrodynamic calculation \cite{Stoecker, Stoecker2013}. Other final-state interaction effects, such as annihilation \cite{Fuqiang} and hadronic potentials \cite{AMPTpotentials}, complicate the simplified picture above. 

%%%%%%%%%%%%%%%%%%%%%%%%%%% FIG 4 %%%%%%%%%%%%%%%%%%%%%
%%%%%%%%%%%%%%%%%%%%%%%%%%%%%%%%%%%%%%%%%%%%%%%%%%%%%%%
Fig.~\ref{fig4}(c) reveals that the $v_1(y)$ slope for net protons is negligibly different from protons at 11.5 and 7.7 GeV, but then crosses zero between 27 GeV and 39 GeV, and remains positive up to 200 GeV.  UrQMD \cite{urqmd} again shows a monotonic trend, with a positive slope at all energies. The observed beam energy of the minimum in $v_1(y)$ slope for both protons and net-protons is higher than the energy of the minimum in the hydrodynamic prediction \cite{Px-v1}. Recent hydrodynamic calculations confirm this prediction, but yield larger $v_1$ magnitudes than observed \cite{Hybrid-v1}. A recent hybrid calculation, featuring Boltzmann transport with an intermediate hydrodynamic stage \cite{Hybrid2008}, does not show a minimum or a sign change in $dv_1/dy$ \cite{Hybrid-v1}.

The beam energy region where we observe the minimum in $v_1(y)$ slope for all protons and net-protons coincides with a high degree of stopping \cite{stopping}. It is not far above the AGS/E895 energy region (lab energies of 2--8 $A$ GeV) where the spectator matter separates from the participants quickly enough so that its influence on the flow in the mid-rapidity zone decreases steeply as the energy is increased further \cite{Pawel,Aihong-v1Model}.  Nuclear transport models like UrQMD ought to clarify whether or not purely hadronic physics could account for the observed minimum, and for the double sign change in the case of net protons.  Further work towards a more complete theoretical understanding of the present observations is needed.  To better understand the possible role and relevance of stopping, measurements as a function of centrality would be helpful, but available event samples are too small for this purpose. We note that the observations in Fig.~\ref{fig4}(b) and (c) qualitatively resemble predicted signatures of a first-order phase transition between hadronic and deconfined matter ~\cite{Rischke,Brachmann,Csernai,Stoecker,Herrmann,Bozek}. 

In summary, we report directed flow for charged pions, protons and antiprotons in $\sqrt{s_{NN}} =$ 7.7 - 200 GeV Au+Au collisions in the STAR detector at RHIC.  At intermediate centralities, $dv_1/dy$ near mid-rapidity for pions and antiprotons is negative at all measured energies, while the proton slope changes sign from positive to negative between 7.7 and 11.5 GeV, shows a minimum between 11.5 and 19.6 GeV, and remains small but negative up to 200 GeV. In the same centrality region, the net-proton $v_1(y)$ slope also shows a minimum between 11.5 and 19.6 GeV, and changes sign twice between 7.7 and 39 GeV. These findings are qualitatively different from the predictions of the UrQMD transport model, which exhibits a monotonic trend in the range $\sqrt{s_{NN}} = 7.7 - 200$ GeV.  The observed minimum for protons and net protons resembles the predicted ``softest point collapse" of flow and is a possible signature of a first-order phase transition between hadronic matter and a deconfined phase.  

%%%%%%%%%%%%%%%%%%%%%%%%%%%%%%%%%%%%%%%%%%%%%%%%%%%%%%%%%%%
We thank H. Petersen, H. Steinheimer and H. St\"ocker for helpful discussions. We also thank the RHIC Operations Group and RCF at BNL, the NERSC Center at LBNL, the KISTI Center in Korea and the Open Science Grid consortium for providing resources and support. This work was supported in part by the Offices of NP and HEP within the U.S. DOE Office of Science, the U.S. NSF, CNRS/IN2P3, FAPESP CNPq of Brazil, Ministry of Ed. and Sci. of the Russian Federation, NNSFC, CAS, MoST and MoE of China, the Korean Research Foundation, GA and MSMT of the Czech Republic, FIAS of Germany, DAE, DST, and CSIR of India, National Science Centre of Poland, National Research Foundation (NRF-2012004024), Ministry of Sci., Ed. and Sports of the Rep. of Croatia, and RosAtom of Russia.

%\end{linenumbers}


\begin{thebibliography}{99}
\bibitem{simulations}
This effect has been estimated using GENBOD \cite{GENBOD} and MEVSIM \cite{MEVSIM} event generators.
\end{thebibliography}
\end{document}